\documentclass[final,1p,times]{elsarticle} 

\usepackage{graphicx}
\usepackage{amssymb} 
\usepackage{amsthm} 
\usepackage{lineno}

\newcommand{\pt}{$p_{T}$}
\newcommand{\kt}{$k_{T}$}

\def\be{\begin{equation}}
\def\ee{\end{equation}}
\def\bea{\begin{eqnarray}}
\def\eea{\end{eqnarray}}

\journal{Nuclear Physics A} 

\begin{document}

\begin{frontmatter} 

\title{Review of Recent Developments in the CGC}

\author{Adrian Dumitru}
\address{RIKEN BNL Research Center, Brookhaven National
  Laboratory, Upton, NY 11973, USA\\
Department of Natural Sciences, Baruch College, CUNY,\\
17 Lexington Avenue, New York, NY 10010, USA}


\begin{abstract} 
I review recent developments in the CGC approach to high-energy
collisions. The focus is on topics related to the Quark Matter
conference, specifically on predictions for the p+Pb run at
the LHC; as an added bonus some of these predictions are confronted
with first data taken during the p+Pb machine performance test
at the LHC performed in september 2012. I also highlight recent work
related to particle production fluctuations, suppression of nuclear
modification factors in p+A from rcBK evolution, (partial) NLO
corrections in forward production, and photon-hadron as well as
di-hadron correlations.
\end{abstract} 

\end{frontmatter} 


\section{Introduction}

In the Bjorken-Feynman parton model the constituents of a boosted
hadron are viewed as dilute non-interacting quanta collinear to the
beam. This picture has had tremendous success describing the proton
structure at very short distances, high-$E_T$ jets in hadronic
collisions, and more. However, for a probe of fixed size $r$ (well
below the confinement scale of $\sim 1$~fm) in the high-energy limit
this picture is expected to break down. This is due to ample radiation
in QCD of soft gluons as phase space opens up~\cite{GLR}. As a consequence,
although the QCD coupling at the scale $r$ may be small there are many
${\cal O}(\alpha_s^{-1})$ partners to interact with. Thus, at high
energies hadrons rather correspond to a dense system of gluons and
non-perturbatively strong fields,
$A^\mu\sim1/g$~\cite{Mueller:1999wm}. Thanks to the high occupation
number these can be described as classical color fields sourced by the
``valence'' charges at higher rapidities which have been integrated
out~\cite{MV}. The Color Glass Condensate (CGC) is an effective theory
describing the dynamics of non-linear color fields at high
energies. Exploring its properties and systematically improving its
accuracy is one of the main goals of the high-energy nuclear physics
programme.

At present, the most suitable process for studies of high gluon
density QCD is p+A collisions where a relatively dilute projectile
probes a dense target. Particle production in the forward region of
d+Au collisions at RHIC energy has discovered an interesting
suppression relative to a simple superposition of binary p+p
collisions~\cite{Arsene:2004ux}. However, due to limited phase space
those measurements are restricted to relatively low transverse momenta
and can not probe accurately the tails of the intrinsic gluon
transverse momentum distributions. Also, quantitative calculations
suffer from large uncertainties from a variety of sources: the
large-$x$ parton distributions of the projectile ($x\sim 0.1 - 0.8$),
the large-$z$ behavior of parton fragmentation functions ($z>0.6$),
NLO corrections to the ``hybrid formalism'' at high {\pt} (see below),
and so on. Presently, many eyes in this community are therefore
focused on the upcoming p+Pb run at the LHC at $\surd s=5$~TeV which
will open up phase space for small-$x$ physics tremendously. This
write-up shall therefore also focus mainly (but not exclusively) on
recent predictions for p+Pb collisions at the LHC based on the CGC
approach.

The frozen valence charge density per unit transverse area, $\rho$, in
a hadron or nucleus is a random variable with a distribution
determined by an effective action $S[\rho]$. In the limit of a large
number of valence sources this distribution is Gaussian~\cite{MV},
\be \label{eq:S_MV}
S_{\rm MV}[\rho] = \int d^2r \, dy \, \frac{\rho^2}{2\mu^2}~.
\ee
The variance $\mu^2$ is proportional to the thickness of the target
($=A^{1/3}$ for a nucleus, on average over impact parameters) since
the sources act coherently. The classical field $A^\mu_{\rm cl}$ is
obtained from $\rho$ by solving the Yang-Mills equations, analogous to
the Weizs\"acker-Williams approach in electrodynamics.

At the purely classical level the saturation momentum is energy
independent. Energy dependence arises from quantum corrections to the
classical field. It is not presently known how to incorporate all
quantum corrections but a specific class of quantum fluctuations,
those which are approximately boost invariant and are proportional to
$\alpha_s \log 1/x$ can be re-summed. For the two-point correlation
function of light-like Wilson line operators, the so-called dipole
scattering amplitude
\be
\mathcal{N}(r) = \frac{1}{N_c}\left< {\rm tr}~V(0)\, V^\dagger({\bf
  r}) \right>~,
\ee
this is accomplished by the
Balitsky-Kovchegov (BK) equation:
\be
  \frac{\partial\mathcal{N}(r,x)}{\partial\ln(x_0/x)}=\int d^2 r_1\
  K(r,r_1,r_2) \left[\mathcal{N}(r_1,x)+\mathcal{N}(r_2,x)
-\mathcal{N}(r,x)-\mathcal{N}(r_1,x)\,\mathcal{N}(r_2,x)\right]~.
\label{eq:BK}
\ee
Due to the presence of non-linear effects this equation leads to {\em
  saturation} of the scattering amplitude at large dipole sizes $r$ or
small intrinsic transverse momenta \kt. Evolution thus generates a
dynamical scale $Q_s(x)$ which grows with energy (or $1/x$). In
practice, eq.~(\ref{eq:BK}) is nowadays solved with running coupling
accuracy~\cite{rcBK}, and one of the main goals in this field is to
test whether the evolution speed predicted by rcBK is consistent with
observations or if improved accuracy is required.  From the Fourier
transform of $\mathcal{N}(r,x)$ (transformed to the adjoint
representation) one obtains the (dipole) unintegrated
gluon distribution (UGD) $\Phi(k_\perp,x)\sim k_\perp^2 \;
\mathcal{N}_A(k_\perp,x) / \alpha_s(k_\perp^2)$.

Solving~(\ref{eq:BK}) requires an initial condition
$\mathcal{N}(r,x_0)$, where the reference rapidity $\log 1/x_0$ is
typically estimated to be $Y_0 = \log 1/x_0\simeq \log 100 =
4.6$. $\mathcal{N}(r,x_0)$ is the dipole scattering
amplitude corresponding to the initial classical field at the
reference rapidity $Y_0$. In the MV model~(\ref{eq:S_MV}) it is given by
\be
\mathcal{N}(r,x_0) = 1 - \exp \left[- \frac{1}{4}
\left(r^2 Q_s^2(x_0) \right)^\gamma \log \frac{1}{r \Lambda}
  \right]~~~~\mbox{(MV: $\gamma=1$; ~AAMQS: $\gamma\simeq1.1$)},  
\label{eq:N0}
\ee
with $Q_s^2(x_0) \sim g^4\mu^2$ and $\gamma=1$. Ref.~\cite{AAMQS}
performed detailed fits to (the most recent) HERA DIS data and found
that a larger value for the initial ``anomalous dimension'' is
preferred. Below, we shall see that semi-hard {\pt} distributions in
p+p collisions are consistent with the AAMQS initial condition coupled
with rcBK evolution but clearly exclude the MV model initial condition
for protons. As a first quantitative achievement we thus note that the
proton rcBK-UGD at small $x<x_0$ (averaged over impact parameter) is
now known to some degree of accuracy. However, the physical origin of
the AAMQS correction to $\gamma$, which corresponds to a suppression
of the classical $\Phi(k_\perp,x)\sim k_\perp^{-2}$ bremsstrahlung
tail, is not very well understood. Since the valence charge density in
a proton at $x_0\simeq 0.01$ is not very large one may consider
corrections of order $\sim\rho^4$ to the MV model
action~\cite{Dumitru:2011ax}.

\section{Multiplicities and multiplicity distributions}
The \pt-integrated multiplicity of charged particles is the most basic
observable in particle collisions. For a variety of reasons though,
the ``details'' are too many to discuss here, it is also hard to compute
precisely, unfortunately. That said, one may hope that the main
dependence of $dN/d\eta$ on energy and system size is through the
saturation scale $Q_s$ and so it is certainly warranted to check how
this compares to data.

\begin{figure}[htbp]
\begin{center}
\includegraphics[width=0.6\textwidth]{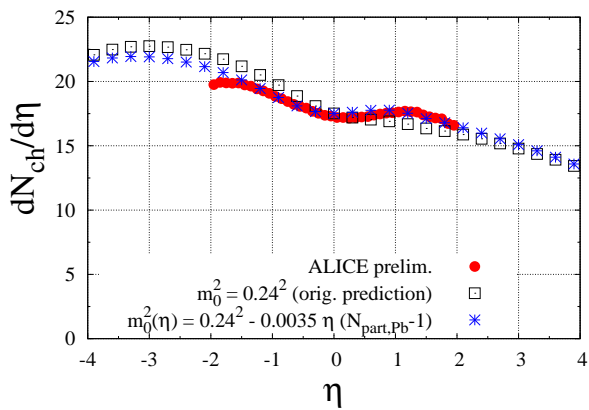}
\end{center}
\vspace*{-0.5cm}
\caption{Charged particle rapidity distribution at 5~TeV from the KLN
  gluon saturation model~\cite{Dumitru:2011wq}.}
\label{fig:KLN_pPb5000}
\end{figure}
This issue was first addressed by the KLN model which indeed is in
reasonable agreement, certainly well within its level of accuracy and
credibility, with the centrality dependence of the multiplicity at
midrapidity in Au+Au and d+Au collisions at RHIC, Pb+Pb collisions at
LHC, and even p+p collisions from 0.9 -- 7~TeV; for the most recent
compilation and references to the original KLN papers we refer to
ref.~\cite{Dumitru:2011wq}. That paper also presented a prediction for
p+Pb collisions at 4.4~TeV; an update for 5~TeV is shown in
fig.~\ref{fig:KLN_pPb5000}. Other predictions for the
multiplicity in p+Pb collisions at the LHC which instead use rcBK or
IPsat UGDs, differ in their hadronization prescription or in the
treatment of the nuclear geometry and of Glauber fluctuations, lead to
very similar predictions at $\eta=0$ (deviations from the central
value are
$\pm15\%$)~\cite{Tribedy:2011aa,Rezaeian:2011ia,Albacete:2012xq}. This
confirms that the energy dependence of particle production is
determined mainly by the growth of the saturation scale. It would be
interesting also to compare theory and data for different cuts on
$N_{\rm part}$. The prediction is in very good agreement with
preliminary ALICE data~\cite{ALICE_pPbdNchdeta} at midrapidity, with a
slightly too steep $\eta$-dependence.

\begin{figure}[htbp]
\begin{center}
\includegraphics[width=0.48\textwidth]{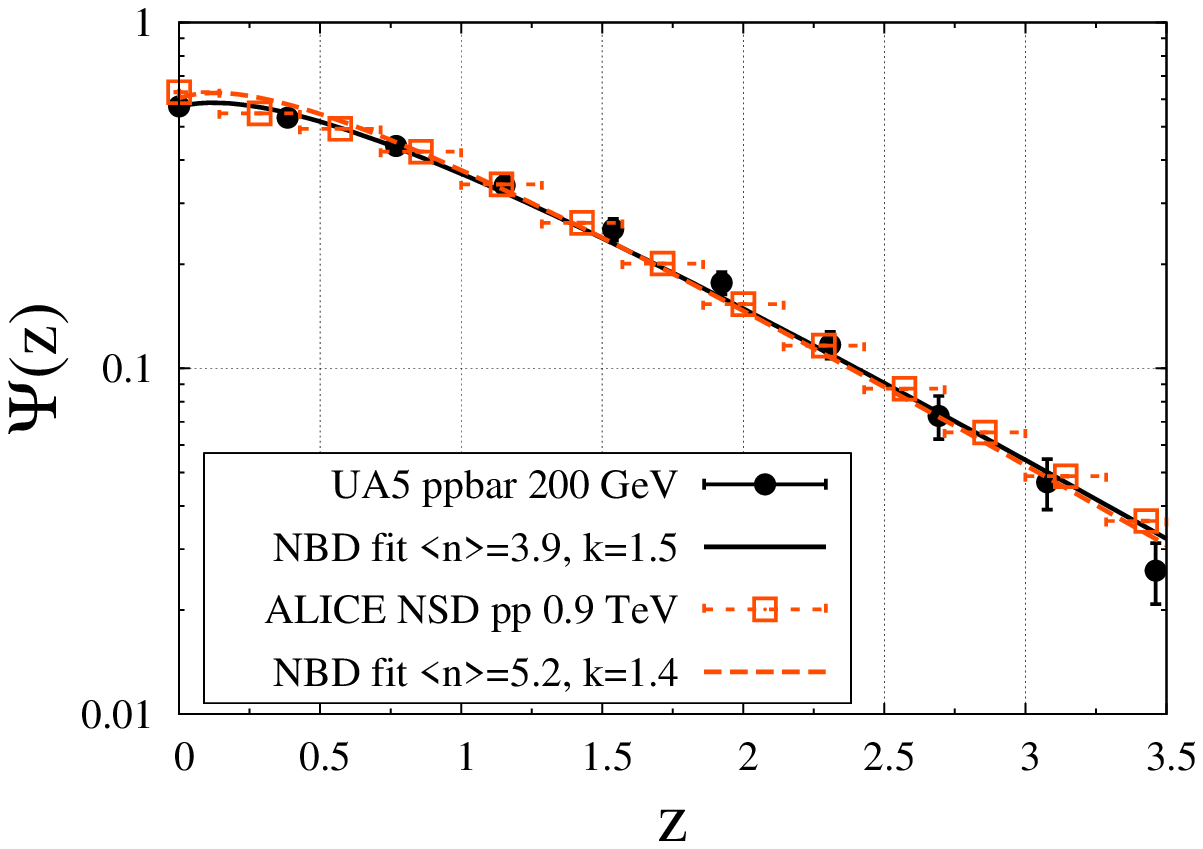}
\includegraphics[width=0.48\textwidth]{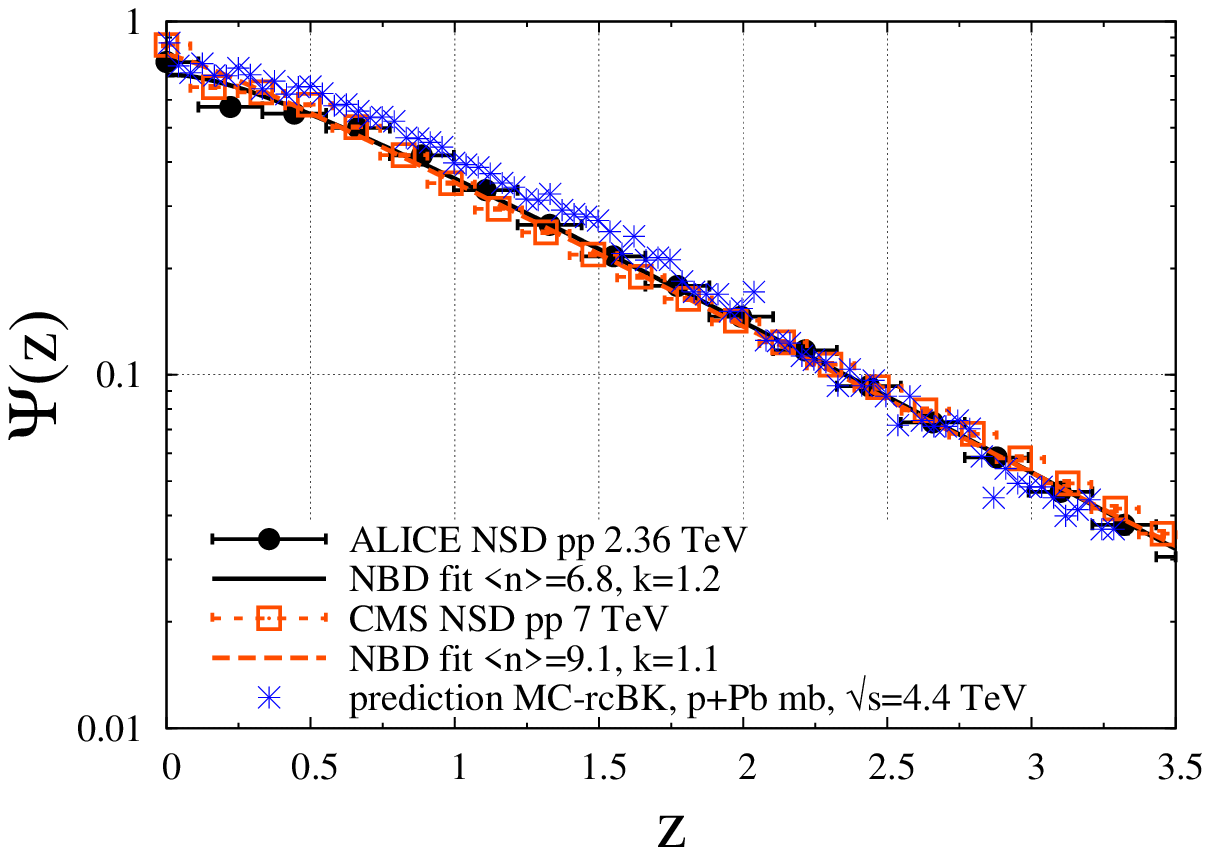}
\end{center}
\vspace*{-0.4cm}
\caption[a]{KNO scaling plot of charged particle multiplicity
  distributions at $|\eta|<0.5$ in NSD collisions at various
  energies~\cite{Ansorge:1988kn} and NBD fits; $z\equiv N_{\rm ch}/\langle
  N_{\rm ch}\rangle$ and $\Psi(z)\equiv \langle N_{\rm ch}\rangle\,
  P(N_{\rm ch})$. Note that the quoted mean multiplicity includes
  neutral particles; also, that here the fluctuation parameter $k$ is
  integrated over the transverse plane of the collision. p+Pb
  prediction from ref.~\cite{Dumitru:2012yr}.}
\label{fig:KNOfits}
\end{figure}
\begin{figure}[htbp]
\begin{center}
\includegraphics[width=0.55\textwidth]{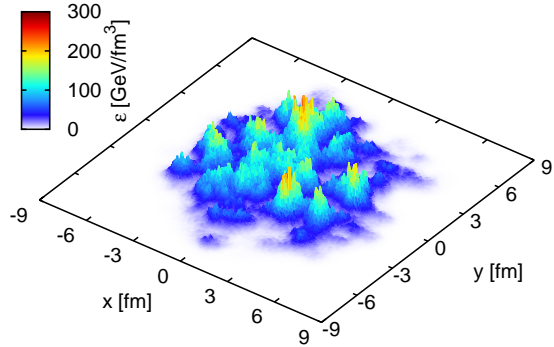}
\end{center}
\vspace*{-0.6cm}
\caption[a]{Fluctuating energy density distribution at midrapidity and time
  $\tau=0.2$~fm/c in an A+A collision~\cite{Schenke:2012wb}.}
\label{fig:fluc_edens}
\end{figure}

One can also study the entire multiplicity distribution rather than
just its mean. Assuming that particle production is due to strong
classical fields, multiplicity fluctuations arise due to color charge
density fluctuations of the effective sources at midrapidity. For a
quadratic action such as in the MV model this leads (approximately) to
a negative binomial distribution~\cite{Gelis:2009wh}; corrections are
illustrated in ref.~\cite{DumitruPetreskaKNO}. At midrapidity the bulk
of the multiplicity distributions from p+p collisions at energies 0.2
-- 7~TeV can indeed be described quite well by
NBDs~\cite{Tribedy:2011aa,Dumitru:2012yr} which in fact exhibit KNO
scaling to a good
approximation~\cite{DumitruPetreskaKNO,Dumitru:2012yr}, see
fig.~\ref{fig:KNOfits}. Approximate KNO scaling has been predicted to
persist even for minimum-bias p+Pb collisions at the
LHC~\cite{Dumitru:2012yr}. This represents an important check of our
current understanding of multi-particle production. Most importantly
such {\it intrinsic} particle production fluctuations affect the
initial state for hydrodynamics in A+A
collisions~\cite{Dumitru:2012yr,Schenke:2012wb} and thus could
manifest in final-state flow and angular correlations. In the CGC
approach, they are expected to occur on sub-nucleon distance scales on
the order of $\sim1/Q_s$~\cite{Schenke:2012wb}, as shown in
fig.~\ref{fig:fluc_edens}.  In this context it is interesting to note
that CGC initial state models which do {\em not} incorporate intrinsic
particle production fluctuations~\cite{DrescherNara} appear to be
inconsistent with the distributions of angular flow harmonics
presented at this conference by the ATLAS collaboration~\cite{Jia}.

\section{Nuclear modification factor $R_{\rm pA}(p_T)$}

Transverse momentum distributions of produced particles provide more
detailed information than \pt-integrated multiplicities. In
particular, they can probe the tail of the distribution of gluon
intrinsic \kt. To compare to small-$x$ QCD evolution one should
restrict to $p_T/(\langle z\rangle \surd s)\, e^{\pm y} <x_0 \sim 0.01$
where $\langle z\rangle$ is the typical momentum fraction in fragmentation.

\begin{figure}[htbp]
\begin{center}
\includegraphics[width=0.48\textwidth]{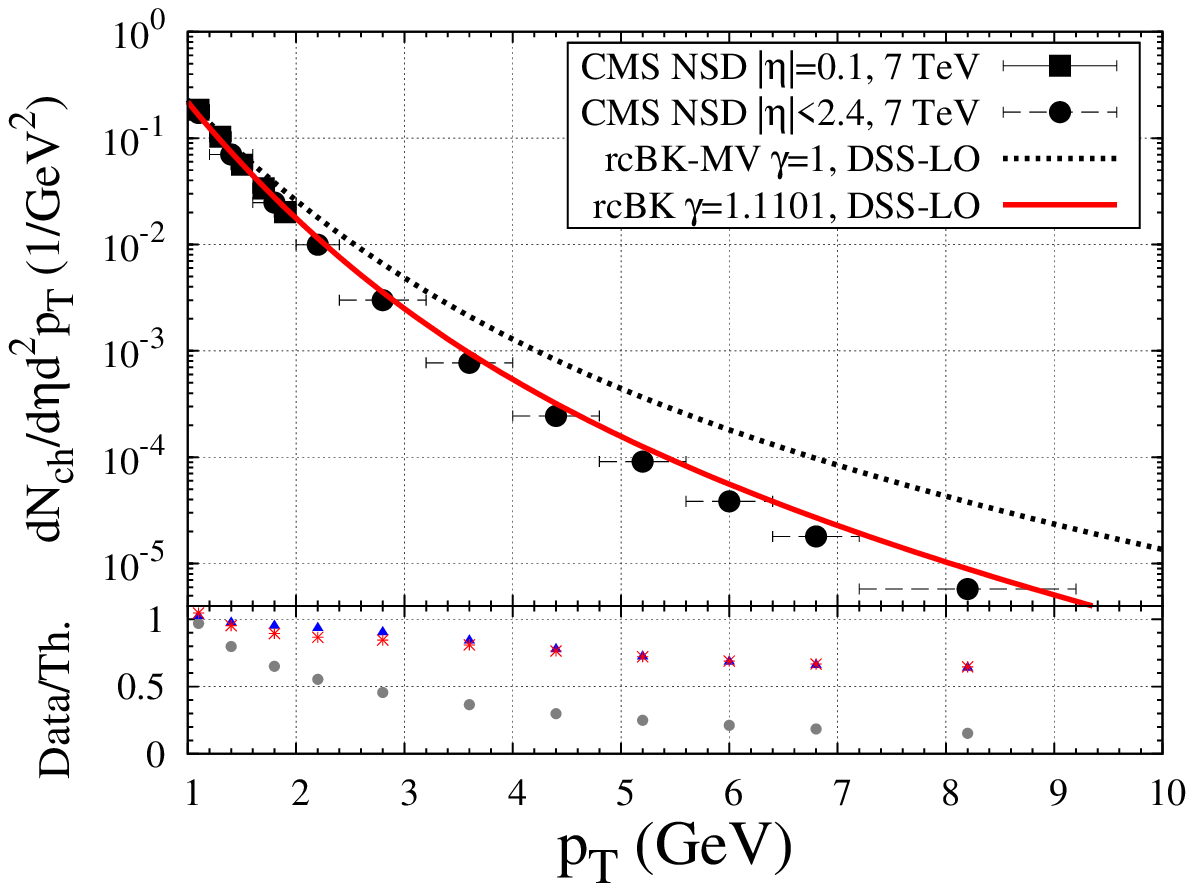}
\includegraphics[width=0.48\textwidth]{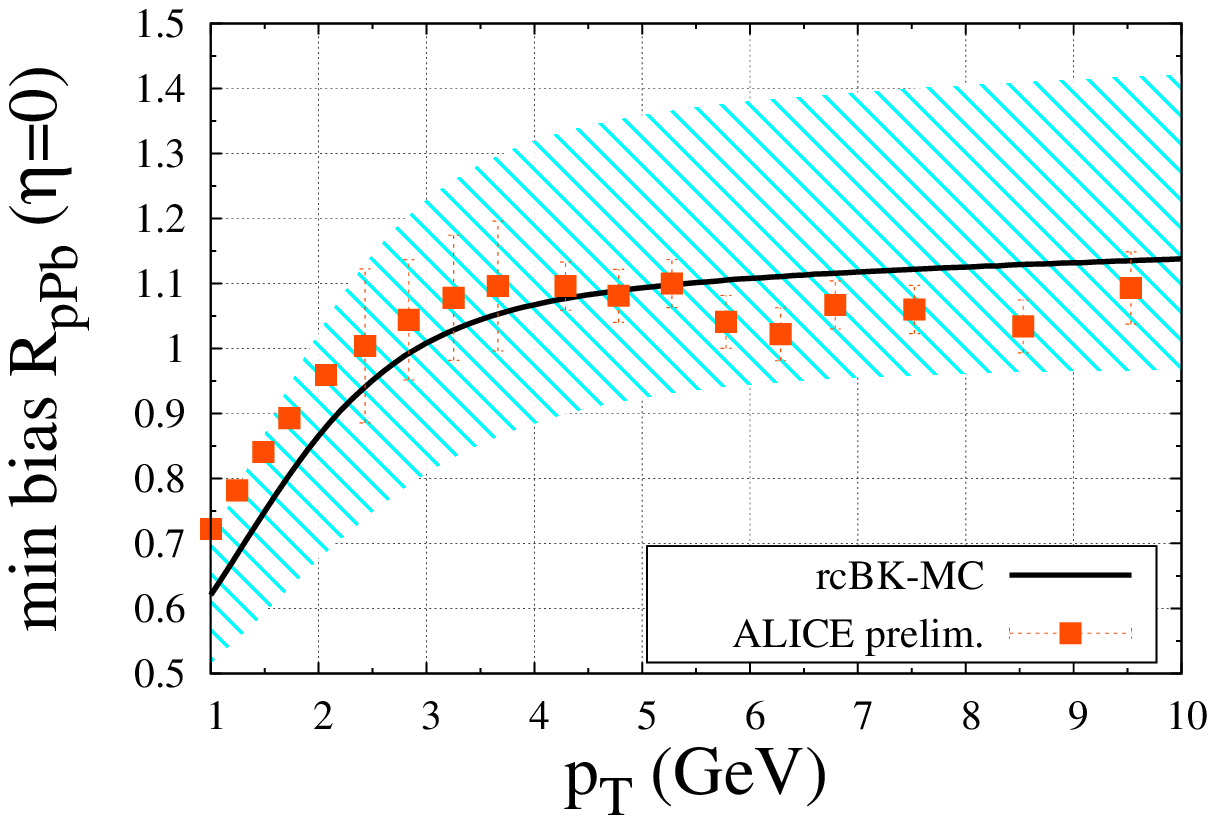}
\end{center}
\vspace*{-.4cm}
\caption{Left: Single-inclusive \pt-distribution in p+p collisions at
  7~TeV~\cite{Albacete:2012xq}; the curves show results obtained with
  rcBK evolved UGDs starting from two different initial conditions at
  $x_0=0.01$. Theory $K$-factor fixed at \pt = 1~GeV.
  Right: nuclear modification factor $R_{\rm pPb}(p_T)$ at $\eta=0$
  (w/o rapidity shift)~\cite{Albacete:2012xq}.
}
\label{fig:ppdNdpt7000}
\end{figure}
As a first step, it is important to check the spectrum in p+p
collisions to establish consistency with the UGD fitted to HERA DIS
data. Furthermore, a correct p+p limit is required for trustworthy
minimum-bias p+A spectra and for $R_{\rm pA}$ nuclear modification
ratios. The result~\cite{Albacete:2012xq} is shown in
fig.~\ref{fig:ppdNdpt7000} for two different initial
conditions. Clearly, the MV model initial condition with rcBK
evolution is excluded by the LHC data which exhibits strong
suppression of the high-\kt tails as compared to this UGD. On the
other hand the AAMQS initial condition which has been carefully fitted
to HERA data also agrees reasonably well with the CMS
spectrum~\cite{Khachatryan:2010us} and thus conforms to process
independence of the (dipole) UGD.

Fig.~\ref{fig:ppdNdpt7000} also shows a prediction for the nuclear
modification factor for p+Pb collisions at 5~TeV. The prediction and
its many uncertainties are discussed in detail in
ref.~\cite{Albacete:2012xq}. Generically, for all UGDs there is a
suppression of particle production at $p_T=1$~GeV which grows stronger
for more ``central'' collisions (higher $N_{\rm part}$). $R_{\rm pPb}$
then approaches and may even exceed unity at higher \pt, as the
evolution window shrinks; recall that in hadronic collisions, at fixed
rapidity $x \propto p_T$. Also, fluctuations in the thickness of the
target amplify higher-twist ``anti-shadowing'' effects. The shape of
the $R_{\rm pPb}(p_T)$ curves provides a test for the evolution speed
predicted by rcBK; again, within uncertainties the prediction matches
the preliminary data from ALICE.

A very important recent development addresses corrections to the
so-called ``hybrid formalism'' for particle production in asymmetric
kinematic configurations (such as forward production in p+A
collisions): large-$x$ parton distributions are described within DGLAP
while the small-$x$ target field is obtained from rcBK, for
example. Until recently only the LO expression corresponding to
elastic scattering of collinear projectile partons on the dense target
field was known~\cite{Dumitru:2005gt}. Ref.~\cite{Altinoluk:2011qy}
computed inelastic corrections corresponding to large-angle emission
of a gluon in the projectile wave function which then exchanges little
transverse momentum with the target. This correction is formally
suppressed by one power of $\alpha_s$ but enhanced by $\log
p_T/Q_s$. Numerical
evaluations~\cite{Albacete:2012xq,JalilianMarian:2011dt} show that
at high {\pt} inelastic corrections drive $R_{\rm pA}$
up. However, for realistic values of $\alpha_s$ this contribution can
be very large, especially for p+p collisions where $Q_s$ is relatively
small~\cite{Albacete:2012xq}. The full NLO expression has been derived
recently~\cite{Chirilli:2011km} and awaits numerical evaluation; this
is crucial in order to understand the accuracy with which we are able
to compute particle production in the forward region.

\begin{figure}[htbp]
\begin{center}
\includegraphics[width=0.49\textwidth]{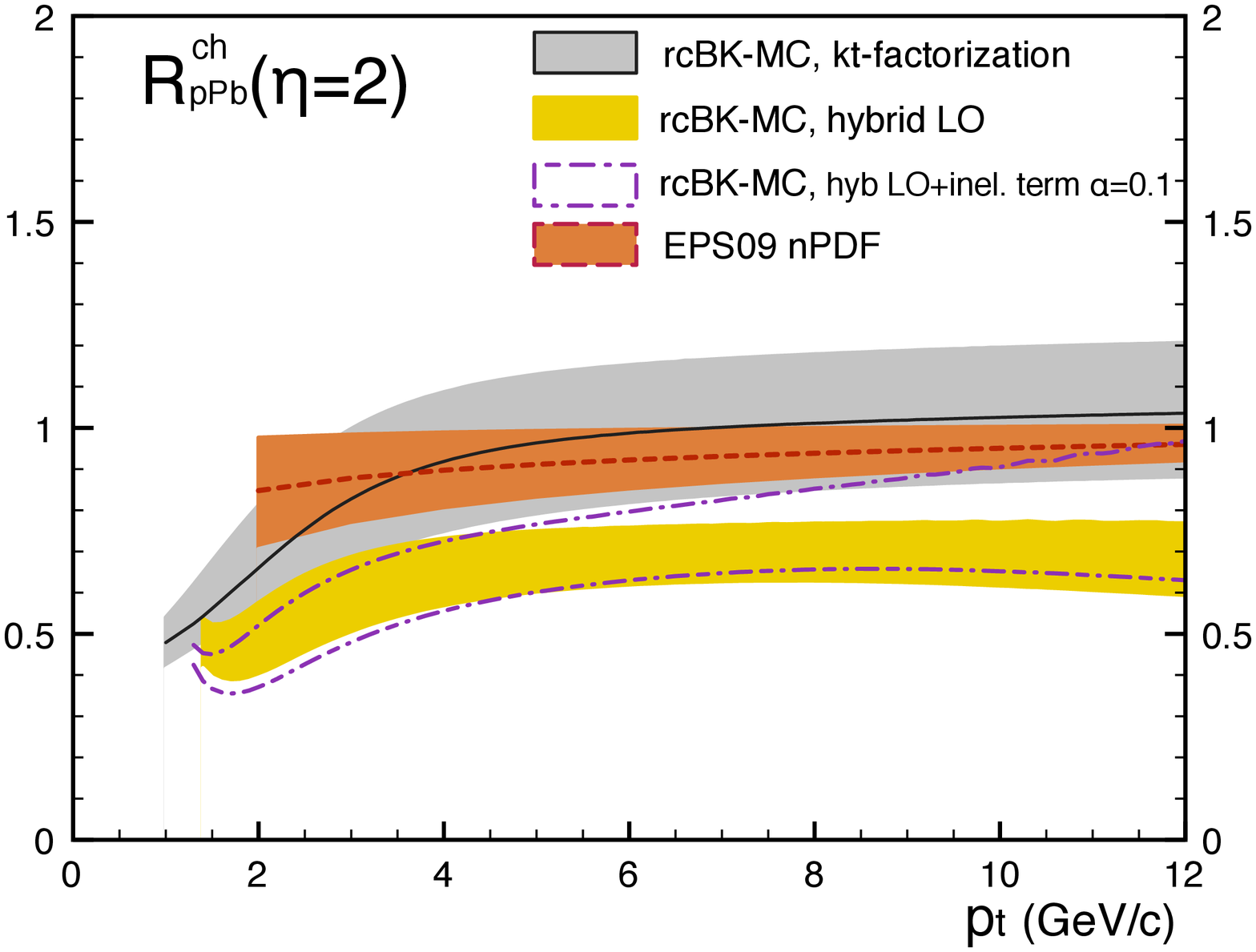}
\includegraphics[width=0.49\textwidth]{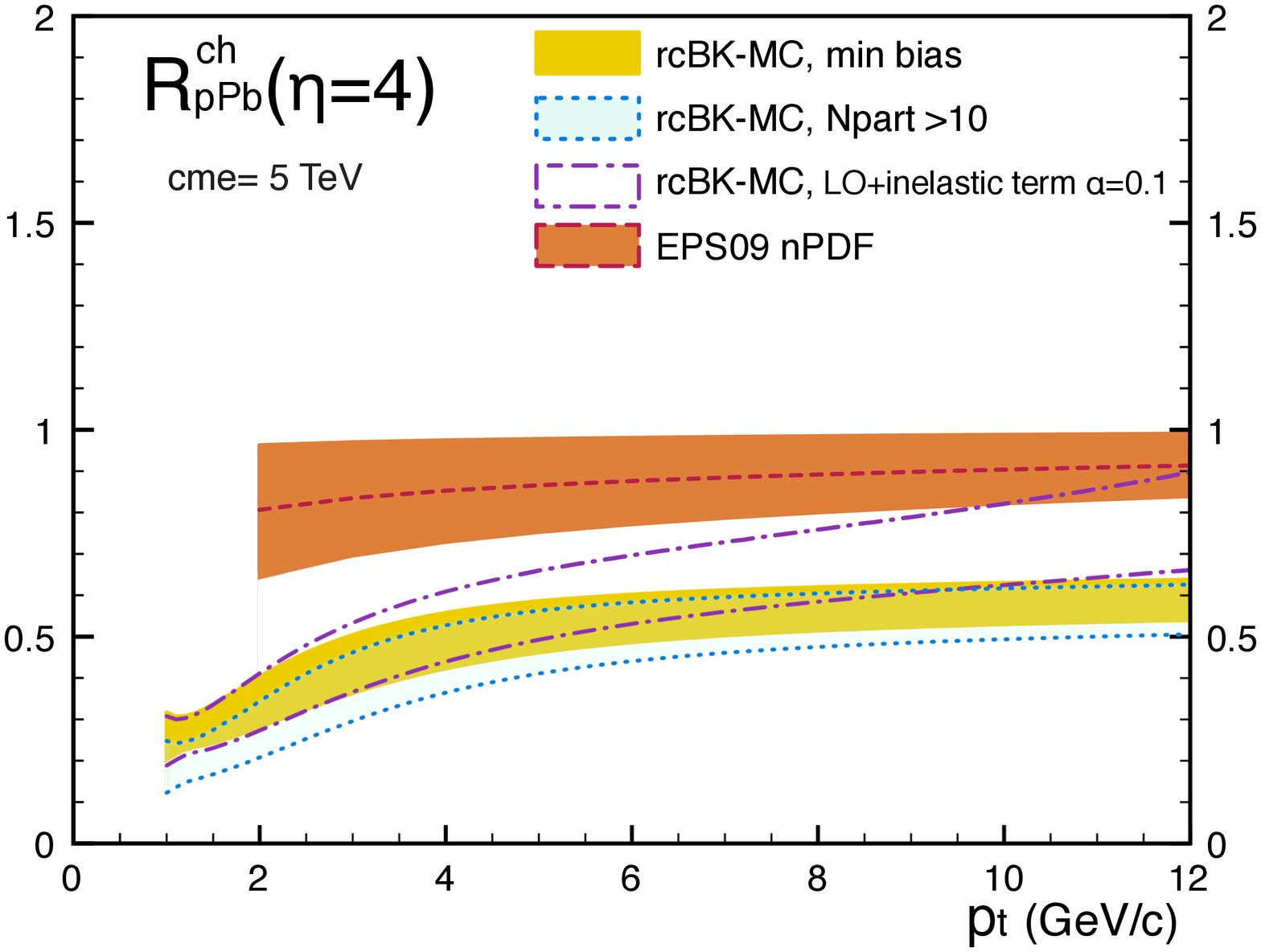}
\end{center}
\vspace*{-.4cm}
\caption{Nuclear modification factor $R_{\rm pPb}(p_T)$ at $\eta=2$, 4
  obtained from the hybrid formalism~\cite{Albacete:2012xq}.
}
\label{fig:hybrid_forward_pPb}
\end{figure}
A prediction for $R_{\rm pA}$ at forward rapidities is shown in
fig.~\ref{fig:hybrid_forward_pPb}. As expected, generically the suppression
increases with rapidity and could grow much stronger than predicted by
some current leading-twist shadowing
approaches~\cite{Eskola:2009uj,QuirogaArias:2010wh}. Notice the large
correction (mainly to the p+p baseline) at high {\pt} due to the
inelastic term, even for very small $\alpha_s=0.1$; as already
mentioned above more reliable predictions at high {\pt} require full
NLO accuracy.

\section{$\gamma$-hadron and hadron-hadron azimuthal correlations}

There has been a lot of activity recently also to predict azimuthal
correlations which could be a very powerful tool for detecting high
density effects. In essence, $2\to2$ hard scattering should lead to
approximately back to back correlations but if the projectile
parton scatters off a strong target field that correlation should
weaken or even disappear entirely. For the strongest effect,
transverse momenta should be on the order of $Q_s$, not far above.

\begin{figure}[htbp]
\begin{center}
\includegraphics[width=0.46\textwidth]{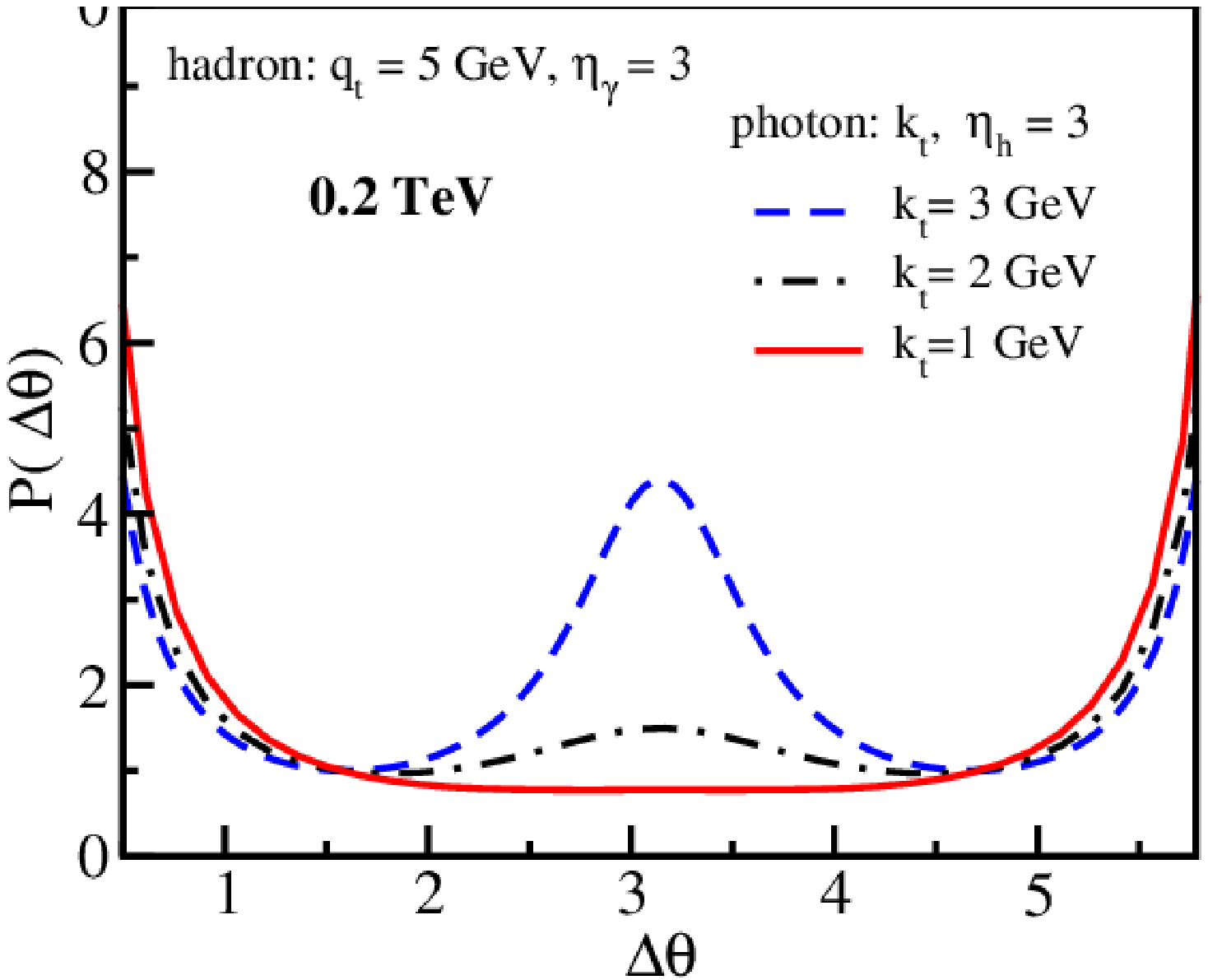}
\includegraphics[width=0.52\textwidth]{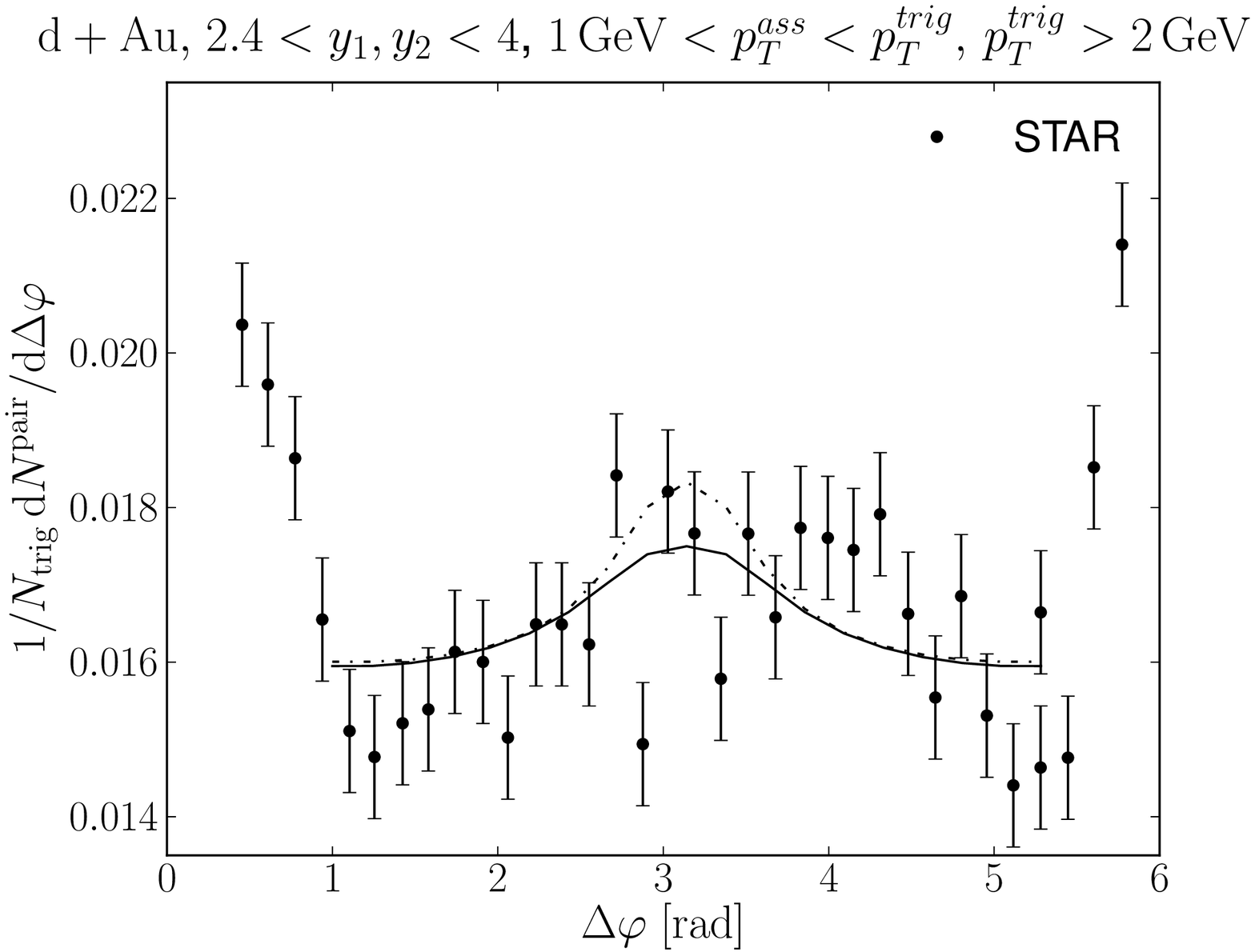}
\end{center}
\vspace*{-.4cm}
\caption{Left: predicted $\gamma$-hadron azimuthal correlations in the
  forward region of d+Au collisions at RHIC~\cite{JalilianMarian:2012bd}.
  Right: di-hadron correlations (incl.\ pedestal from double parton
  scattering) from a Gaussian approximation to JIMWLK
  evolution~\cite{Lappi:2012nh} compared to (preliminary) STAR
  data~\cite{Braidot:2011zj}.
}
\label{fig:correl}
\end{figure}
The most straightforward process for theory is $\gamma$-hadron
correlations. As only one colored particle interacts with the target,
this process, within the hybrid formalism, involves only
the DGLAP PDF of the projectile and the dipole unintegrated gluon
distribution mentioned above (which can be obtained from
rcBK). Theoretical uncertainties should therefore be reasonably small,
allowing for quantitative tests of the theory. A first prediction for
$\gamma$-hadron correlations in the forward region of d+Au collisions
at RHIC is shown in fig.~\ref{fig:correl} \cite{JalilianMarian:2012bd}.

Di-hadron correlations in the forward region are more intricate as now
two colored particles interact with the target. Aside from the BK
dipole the cross section for this process also involves a quadrupole
correlation function~\cite{Marquet:2007vb},
\be
\mathcal{Q} = \frac{1}{N_c}\left< {\rm tr}~V(0)\, V^\dagger({\bf
  r})\, V({\bf u}) \, V^\dagger({\bf v}) \right>~;
\ee
at high \pt, the cross section can be expressed in terms of the dipole
and the (distinct) Weizs\"acker-Williams (WW)
UGDs~\cite{Dominguez:2011wm}. The evolution of the quadrupole with
energy follows from the JIMWLK functional renormalization group
equation. A numerical solution has been presented in
ref.~\cite{Dumitru:2011vk} and turned out to agree rather well, at
least for a few simple configurations, with a Gaussian approximation
(worked out in~\cite{Dominguez:2011wm}) which expresses it as a
(complicated) function of the dipole. A Gaussian approximation is
extremely useful in practice since numerical solutions of the rcBK
equation are rather straightforward; and because the dipole
(incl.\ its initial condition) are rather well constrained by data,
see above. The interest in correlations has also initiated new
theoretical insight in form of a Gaussian approximation to JIMWLK
evolution~\cite{Iancu:2011ns}.

To compare to data, ref.~\cite{Lappi:2012nh} also evaluates the
pedestal due to double parton scattering. The authors observe
broadening of the away side peak, fig.~\ref{fig:correl}, due to the
presence of non-linear effects. An earlier analysis using model
parametrizations for the dipole and WW gluon distributions was
presented in ref.~\cite{Stasto:2011ru}.

\section{Summary}

Tremendous progress has been made in recent years to develop the CGC
effective theory into a quantitative framework which can confront data
from RHIC and LHC. This includes (but is not limited to) improved
computations of multiplicities for a variety of systems and energies;
investigations of multiplicity distributions, KNO scaling in p+p, p+A and
initial-state fluctuations in A+A collisions; semi-hard transverse
momentum distributions and nuclear modification factors with rcBK UGDs
and including Glauber fluctuations; photon production, $\gamma$-hadron
and di-hadron correlations. Currently, the CGC is the only formalism
to address such a broad range of observables systematically and with
some success.

In the near future we should see further improvements of the accuracy
and reliability of some of the predictions, for example a first
computation of particle production in the forward region of p+p and
p+A collisions at full NLO level. Moreover, there is great potential
in understanding initial-state fluctuations in the ``little bang'',
such as the scale on which they occur and how they reflect in the
final state. Further, with respect to gluon correlations in the
boosted wave functions which affect multi-hadron correlations. We are
also presently lacking a description of the transition from semi-hard
{\pt} at small $x$ to the DGLAP regime. These (and other) questions
remain to be solved to achieve a more complete understanding of QCD,
and the structure of matter, at short distances and high energies.

\section*{Acknowledgments}
I thank Quark Matter 2012 for the opportunity to present recent
developments in this field. I also thank all of my collaborators for
their contributions and B.~Schenke, H.~Mantysaari for providing custom
versions of the plots shown in figs.~\ref{fig:fluc_edens} and
\ref{fig:correl}, respectively. This work is supported by the DOE
Office of Nuclear Physics through Grant No.\ DE-FG02-09ER41620 and by
The City University of New York through the PSC-CUNY Research Award
Program, grant 65041-0043.


\end{document}